\documentclass[%
reprint,
superscriptaddress,
 amsmath,amssymb,
prb,
]{revtex4-2}

\usepackage{graphicx}
\usepackage{dcolumn}
\usepackage{bm}

\usepackage{xcolor}

\usepackage[normalem]{ulem}

\newcommand{\kl}[1]{\ensuremath{ \left( #1 \right) }}
\DeclareMathSymbol{\shortminus}{\mathbin}{AMSa}{"39}
\usepackage[pagebackref=false]{hyperref}
\usepackage[capitalize]{cleveref}

\begin{document}

\author{A.~Rothstein}
\email{alexander.rothstein@rwth-aachen.de}
\affiliation{JARA-FIT and 2nd Institute of Physics, RWTH Aachen University, 52074 Aachen, Germany}%
\affiliation{Peter Gr\"unberg Institute  (PGI-9), Forschungszentrum J\"ulich GmbH, 52425 J\"ulich,~Germany}

\author{C. Schattauer}
\affiliation{Institute for Theoretical Physics, TU Wien, 1040 Vienna, Austria}

\author{R. J.~Dolleman}
\affiliation{JARA-FIT and 2nd Institute of Physics, RWTH Aachen University, 52074 Aachen, Germany}%

\author{S.~Trellenkamp}
\affiliation{Helmholtz Nano Facility, Forschungszentrum J\"ulich GmbH, 52425 J\"ulich,~Germany}

\author{F.~Lentz}
\affiliation{Helmholtz Nano Facility, Forschungszentrum J\"ulich GmbH, 52425 J\"ulich,~Germany}

\author{K.~Watanabe}
\affiliation{Research Center for Electronic and Optical Materials, National Institute for Materials Science, 1-1 Namiki, Tsukuba 305-0044, Japan}

\author{T.~Taniguchi}
\affiliation{Research Center for Materials Nanoarchitectonics, National Institute for Materials Science,  1-1 Namiki, Tsukuba 305-0044, Japan}%

\author{D. M. Kennes}
\affiliation{Institute for Theory of Statistical Physics, RWTH Aachen University, and JARA Fundamentals of Future Information Technology, 52062 Aachen, Germany}
\affiliation{Max Planck Institute for the Structure and Dynamics of Matter, Center for Free Electron Laser Science, Hamburg, Germany}

\author{B.~Beschoten}
\affiliation{JARA-FIT and 2nd Institute of Physics, RWTH Aachen University, 52074 Aachen, Germany}%

\author{C.~Stampfer}
\email{stampfer@physik.rwth-aachen.de}
\affiliation{JARA-FIT and 2nd Institute of Physics, RWTH Aachen University, 52074 Aachen, Germany}%
\affiliation{Peter Gr\"unberg Institute  (PGI-9), Forschungszentrum J\"ulich GmbH, 52425 J\"ulich,~Germany}%

\author{F.~Libisch}
\email{florian.libisch@tuwien.ac.at}
\affiliation{Institute for Theoretical Physics, TU Wien, 1040 Vienna, Austria}

\title{Band gap formation in commensurate twisted bilayer graphene/hBN moir\'e lattices}

\date{\today}


\begin{abstract} 
We report on the investigation of periodic superstructures in twisted bilayer graphene (tBLG) van-der-Waals heterostructures, where one of the graphene layers is aligned to hexagonal boron nitride (hBN).
Our theoretical simulations reveal that if the ratio of the resulting two moiré unit cell areas is a simple fraction, the graphene/hBN moir\'e lattice acts as a staggered potential, breaking the degeneracy between tBLG AA sites.
This leads to additional band gaps at energies where a subset of tBLG AA sites is fully occupied. 
These gaps manifest as Landau fans in magnetotransport, which we experimentally observe in an aligned tBLG/hBN heterostructure. 
Our study demonstrates the identification of commensurate tBLG/hBN van-der-Waals heterostructures by magnetotransport, highlights the persistence of moir\'e effects on length scales of tens of nanometers, and represents an interesting step forward in the ongoing effort to realise designed quantum materials with tailored properties.
\end{abstract}

\maketitle
\section{Introduction}
The periodic structure of a crystalline solid is a key factor for its electronic band structure and hence its electrical and optical properties. 
The periodic potential leads to Bloch states as solutions of the Schrödinger equation~\cite{Bloch1929Jul} and to the formation of band gaps -- well-known cornerstones of solid state physics.
In conventional crystalline solids, the modulation of the periodic potential is fixed at the atomic scale. However, progress has been made to engineer artificial lattices -- both at the atomic and nanoscale -- to control electronic band structures, topological material properties, and exotic quantum phases in solids~\cite{Slot2017Jul,Drost2017Jul,Figgins2019Dec,Khajetoorians2019Dec,Lagoin2023Feb,Wang2023Mar}.
A remarkable class of such engineered solids are two-dimensional (2D) van-der-Waals materials with moir\'e lattices, where a small lattice misalignment or twist angle between two layers results in particularly large unit cells -- orders of magnitude larger than the underlying crystal lattice constant~\cite{Kennes2021Feb}.\\

\begin{figure}[!t]
\centering
\includegraphics[draft=false, keepaspectratio=true,clip,width=1\linewidth]{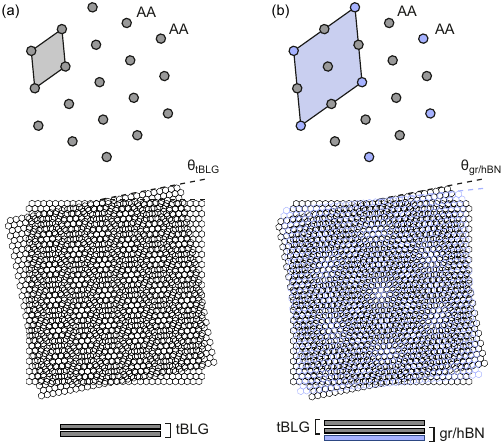}
\caption{
 (a) Schematic representation of a tBLG moir\'{e} lattice. Grey circles represent the AA sites while the gray rhombus denotes the tBLG unit cell. 
 (b) Schematic representation of a composite supermoir\'{e} lattice consisting of a tBLG moir\'e lattice [same as in panel (a)] and a graphene/hBN (gr/hBN) moir\'e lattice. 
 Blue circles represent the AA sites of the resulting supermoir\'e lattice. The shaded rhombus denotes the corresponding supermoir\'e unit cell.
}  
\label{Figure1}
\end{figure}

A prominent example of such a 2D van-der-Waals material 
is twisted bilayer graphene (tBLG) [\cref{Figure1}(a)].
In tBLG, two layers of graphene are stacked with a small %
twist angle, creating a moiré lattice with a large periodic modulation on the order of several nanometers.
The periodic modulation localizes the electronic wavefunction on sites where the carbon atoms of the individual graphene layers align within the trigonal tBLG moir\'e lattice (known as AA sites), leading to the formation of additional band 
gaps~\cite{Cao2016Sep}.
Near the so-called magic angle of $1.1^{\circ}$, tBLG shows a wealth of interesting quantum phases, including superconductivity, correlated insulators and orbital magnetism, all of which depend sensitively on the flat electronic bands in this system~\cite{Cao2018Apr,Cao2018Apr2, Yankowitz2019Mar, Lu2019Oct, Stepanov2020Jul, Saito2020Sep, Zondiner2020Jun, Wong2020Jun, Sharpe2019Aug, Serlin2020Feb, Saito2021Apr, Oh2021Dec, Klein2023Apr}.
Furthermore, tBLG can be combined in a single van-der-Waals heterostructure with a second moir\'e lattice formed by aligning the crystallographic axes of one of the graphene layers with one of the encapsulating hexagonal boron nitride (hBN) layers, exploiting the small lattice constant mismatch between these two materials~\cite{Ponomarenko2013May, Barrier2020Nov, Dean2013May, Hunt2013Jun, Woods2014Jun,Andelkovic2020Feb}.
In this case, an overall \textit{composite} supermoir\'e lattice is formed \cite{Wang2019Dec, Huang2021May, YLi2022}, which breaks the inversion symmetry of tBLG, leading to an altered electronic structure \cite{Long2023Mar, Cea2020Oct, Shi2021Feb, Lin2020Jul, Shin2021Feb}, and to the emergence of novel phases such as ferromagnetism \cite{Serlin2020Feb, Sharpe2019Aug, Sharpe2021May, Bultinck2020Apr, Zhang2019Nov} or anomalous quantum Hall states \cite{Serlin2020Feb, Shi2021Feb}.\\

In this work we present a theoretical model for commensurate tBLG/hBN heterostructures and show that, in the special case where the ratio of the areas of the unit cells of tBLG and of the graphene/hBN (gr/hBN) moir\'e lattice form a simple fraction, the degeneracy of the AA sites of the tBLG is lifted in a periodic manner, which leads to a stronger localization of the electronic wavefunction on a subset of the tBLG AA sites. 
This results in turn in a further flattening of the bands of tBLG and in the appearance of additional band gaps, which manifest themselves as Landau fans at fractional filling factors in the magnetotransport characteristic of the system. 
This prediction is confirmed experimentally by magnetotransport measurements in a tBLG sample aligned to hBN. 
Remarkably, the additional band gaps are not related to full filling of a specific (super)moir\'e cell but to the filling of a \textit{subset} of AA sites of the tBLG, selected by the graphene/hBN moir\'e lattice. \\

\begin{figure*}[!htbp]
\centering
\includegraphics[draft=false, keepaspectratio=true,clip,width=1\linewidth]{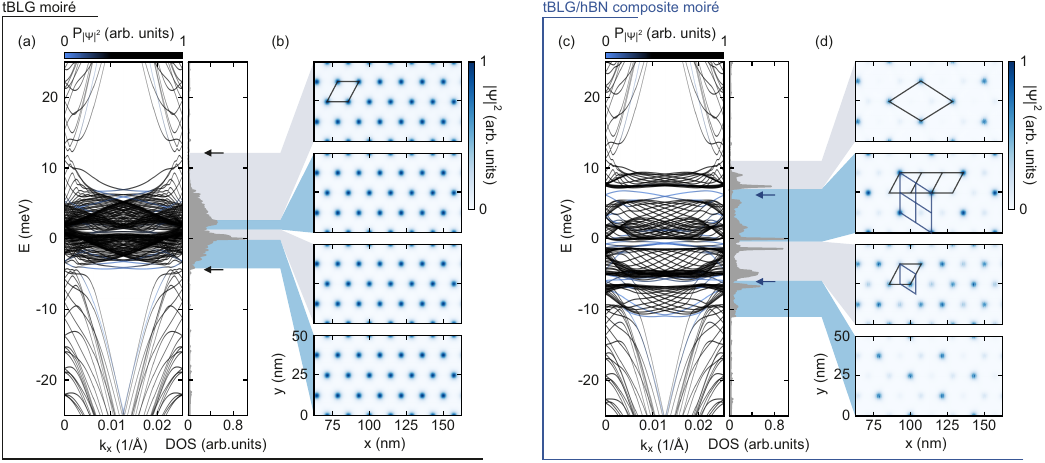}
\caption{
(a) Band structure and density of states calculation of a tBLG moir\'e system at a twist angle of $\theta_\mathrm{tBLG} = 0.987^\circ$ (black arrows mark the tBLG moir\'e-induced single-particle band gaps). 
To distinguish bulk from edge states in our ribbon geometry (width $W\approx 210$ nm), we color each line according to the probability $P_{\vert \Psi \vert^2}$ to find the Bloch state at the center of the ribbon, i.e., within $[0.2, 0.8]\times W$.
(b) Spatial distribution of the charge carrier density in one graphene layer, averaged over different energy windows (shaded areas). 
The charge carrier density accumulates at the AA sites located at the corners of the tBLG moir\'e unit cell (see sketched rhombus) forming an uniform triangular lattice in real space. 
One tBLG moir\'e unit cell contains one AA site (proportional contributions from the corners). 
The $y$-axis is identical for all panels.
(c, d) Same as in (a, b) but with an additional hBN alignment of $\theta_\mathrm{gr/hBN} = 0.62^\circ$. 
Blue arrows mark the positions of the additional gaps forming due to the hBN alignment.
Black rhombus in the upmost panel represents the unit cell of the supermoir\'e lattice defined by the tBLG and the graphene/hBN moir\'e lattices.
The supermoir\'e unit cell contains a total of 3 AA sites.
On this scale three tBLG moir\'e unit cells features the same area as four graphene/hBN moir\'e unit cells (second panel from top).
The presence of the graphene/hBN moir\'e strongly modulates the electron density on the different AA sites of the tBLG moir\'e lattice.
}  
\label{Figure2}
\end{figure*}
\section{Theoretical Model of \lowercase{t}BLG moir\'e systems}
The theoretical modelling of tBLG [see \cref{Figure1}(a)] involves large unit cells -- e.g., about $13 \times 10^3$ atoms for a twist angle of approximately $1^\circ$. 
For tBLG, we parameterize our model directly from density functional theory (DFT), by evaluating the relative, local stacking configurations of the carbon atoms in the top and bottom graphene layers \cite{Fabian2022Oct}.
Using the local DFT parametrization, we assemble a tight-binding Hamiltonian of the entire tBLG moir\'e unit cell. 
Given the different energies of the different stacking configurations, the two layers will stretch and corrugate to minimize their energy. 
We account for this relaxation and for the resulting strain by a membrane model \cite{Nam2017Aug}.
We efficiently describe transport through such a system by 
considering a ribbon geometry of finite width $W$ in one 
direction ($y$-direction)~\cite{Fabian2022Oct}.
We choose  $W \approx 210$ nm (corresponding to 15 tBLG moir\'e unit cells) to reduce finite size effects (for further details see \cref{A5}).
The band structure of the modelled tBLG ribbon features a number of quite flat, nearly-degenerate bands close to the charge neutrality point at $E = 0$~meV [\cref{Figure2}(a)]. 
The corresponding density of states (DOS) shows the expected peak around charge neutrality. 
The energy of the flat bands is confined to a narrow energy interval delimited by the $\nu = \pm 4$ band insulating states [see arrows in \cref{Figure2}(a)] (note, that our single-particle model does not include correlation effects, and therefore no further interaction-induced gaps are visible).
For small fillings, the spatial distributions of the charge carrier density shows the expected localization on the AA sites of the tBLG moir\'e lattice [\cref{Figure2}(b)].
According to this localization on a regular triangular lattice, each tBLG moir\'e unit cell contains a total of four charge carriers due to the spin and valley degeneracy, and a band gap opens at full filling of all AA lattice sites \cite{Bistritzer2011Jul, Cao2016Sep}. \\
\section{Theoretical model of \lowercase{t}BLG/\lowercase{h}BN composite moir\'e systems}
Including into the model the graphene/hBN moir\'e lattice poses a further challenge, as the additional alignment between one graphene layer and the hBN substrate results in an even larger overall composite supermoir\'e unit cell [see \cref{Figure1}(b)].
This applies already to the parametrization by DFT: using the same approach as for tBLG alone would require many additional DFT calculations for all possible shifts between the two graphene layers and all possible alignments of the additional hBN layer. 
To alleviate this further parametrization effort, we only consider the tBLG parameters from DFT, and add the graphene/hBN moir\'e lattice based on an effective potential approach \cite{PhysRevB.90.165404}. 
We use DFT to derive slowly varying, effective potentials that describe the local modulations induced by the graphene/hBN moir\'e lattice in the tBLG. 
Describing the short-range variations induced by the alternating boron and nitrogen atoms requires short-range components of the effective potential, modeled by a second potential with opposite signs on the two sublattices of the graphene layer in contact with the hBN \cite{PhysRevB.89.161401,PhysRevB.84.195414}.
This combination of a smooth and an alternating potential allows us to model the influence of the hBN alignment and to introduce the length scale of the graphene/hBN moir\'e lattice into the tBLG Hamiltonian in an easily adaptable manner.
We have verified numerically that this effective potential approach leads to qualitatively similar predictions (concerning the additional satellite Landau levels, single-particle band gaps etc.) as the full DFT-based atomistic parametrization of the graphene/hBN moir\'e lattice.
While our ribbon geometry allows, in principle, incommensurate unit cell sizes perpendicular to the ribbon, we impose a periodic cell of moderate size in $x$-direction along the ribbon.
For a tBLG twist angle of $\theta_\mathrm{tBLG}= 0.987^{\circ}$ and a graphene/hBN twist angle of $\theta_\mathrm{gr/hBN}= 0.62^{\circ}$, both unit cells feature one identical spatial dimension, which we orient in $x$-direction of our ribbon geometry.
A suitable shift of the hBN layer then allows to obtain a supermoir\'e cell with comparatively small periodicity of $24.8$~nm in $x$-direction. 
This supermoir\'e cell features roughly $8\times 10^4$ carbon atoms, and an area of roughly 300~nm$^2$. 
The 15 tBLG moir\'e unit cells in $y$-direction, perpendicular to the transport direction, correspond to 20 graphene/hBN moir\'e unit cells.
The areas of these two single moir\'e unit cells thus are commensurate, with $A_\mathrm{tBLG}/A_\mathrm{gr/hBN} = 20/15 = 4/3$, i.e. the resulting supermoir\'e unit cell contains three (four) tBLG (graphene/hBN) moir\'e unit cells (see \cref{A5}).\\
\subsection{Influence of the graphene/hBN moir\'e lattice to the tBLG system}
The graphene/hBN moir\'e lattice induces interactions between the tBLG moir\'e flat bands, breaking their degeneracy since adjacent tBLG moir\'e unit cells now  feature different alignments with the graphene/hBN moir\'e lattice.
The resulting level repulsion between the flat bands causes a pronounced broadening of the flat band region [compare \cref{Figure2}(a) and \cref{Figure2}(c)], enhancing the asymmetry of the $\nu = \pm 4$ band gaps.
The DOS for the composite system features a much broader region of peaks, with substantial substructures, delimited by the two band gaps of the bulk tBLG moir\'e lattice excluding the edge states [see \cref{Figure2}(c)].\\

\begin{figure}[!htbp]
\centering
\includegraphics[draft=false,keepaspectratio=true,clip,width=1\linewidth]{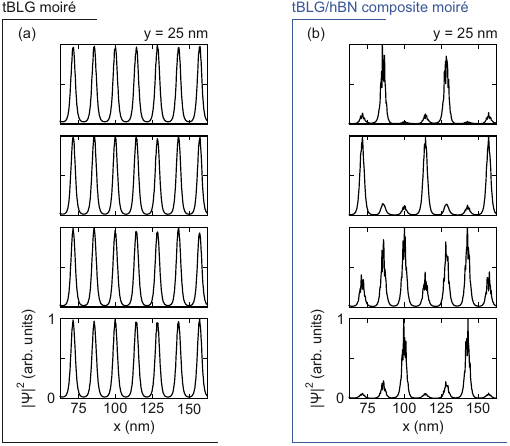}
\caption{ 
(a) Linecuts of the charge carrier density distribution averaged over specific energy ranges as shown in \cref{Figure2}(b) for $y = 25\, \mathrm{nm}$ of the tBLG system.
The energy ranges are given by $E_1 \in [2.5, 12]\, \mathrm{meV}$, $E_2 \in [1.3, 2.5]\, \mathrm{meV}$, $E_3 \in [-0.153, 1.3]\, \mathrm{meV}$ and $E_4 \in [-4.2, -0.153]\, \mathrm{meV}$.
The charge distribution on the different AA sites is uniform for tBLG.
The $y$-axis is identical for all panels. 
(b) Same as in panel (a) but with the additional hBN alignment of $\theta_\mathrm{gr/hBN} = 0.62^\circ$ [linecuts are taken from corresponding panels in \cref{Figure2}(d)].
The energy ranges are given by $E_1 \in [11, 7]\, \mathrm{meV}$, $E_2 \in [-0.4, 7]\, \mathrm{meV}$, $E_3 \in [-6, -0.4]\, \mathrm{meV}$ and $E_4 \in [-11, -6]\, \mathrm{meV}$.
The presence of the graphene/hBN moir\'e lattice strongly modulates the electron density on the tBLG AA sites.} \label{Figure3}
\end{figure}

To further understand the effects of the commensurate graphene/hBN moir\'e lattice, we investigate the spatial distribution of the charge carrier density.  
Because the charge occupation of the triangular lattice of AA sites in the original tBLG is uniform [\cref{Figure2}(b) and linecuts in \cref{Figure3}(a)], the resulting filling factor corresponds to four charge carriers per AA site (due to the spin and valley degree of freedoms). 
The situation is different in the presence of the commensurate graphene/hBN moir\'e lattice [\cref{Figure2}(c,d)]. 
In this case, we observe an additional band gap at an energy of approximately $\left| E \right| \approx \pm 7$~meV [see blue arrows in \cref{Figure2}(c)].
Considering the distribution of charge carrier density $\left|\psi\right|^2$,  we find the occupation of the three AA sites within the supermoir\'e unit cell is now no longer uniform [see shaded regions in \cref{Figure2}(c) and density plots in \cref{Figure2}(d) as well as the corresponding linecuts in \cref{Figure3}(b)]. 
The newly induced energy gap seperates energy regions with varying occupation of the three tBLG AA sites [\cref{Figure2}(d)].
In particular, only one out of the three AA sites is occupied in the range from 7~meV to 11~meV and from -6~meV to -11~meV, [\cref{Figure2}(d) top and bottom pictures].
Indeed, when counting the number of bands above and below the additional gap, we find that one third of the moir\'e flat bands lies above the new gap.
In magnetotransport, this gap should appear as an additional Landau fan at a filling of two out of three states, or $\nu_\mathrm{sat} \approx \pm 4 \times 2/3 = \pm 2.67$. \\

\subsection{Magnetotransport simulation}
\begin{figure}[!]
\centering
\includegraphics[draft=false,keepaspectratio=true,clip,width=1\linewidth]{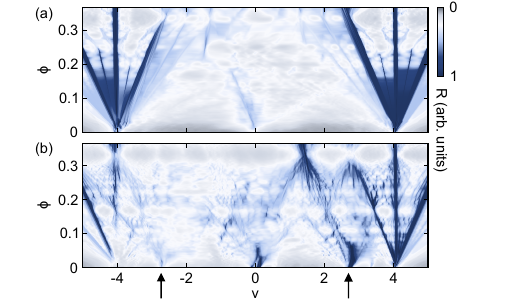}
\caption{ 
(a) Magnetoresistance $R$ [see Eq.~\eqref{eq1}] simulation of a tBLG system with a twist-angle of $\theta_\mathrm{tBLG} = 0.987^\circ$.
Landau levels emerge from the charge neutrality point $(\nu = 0)$ and the single-particle band gaps at full filling of the tBLG moir\'{e} lattice unit cell $(\nu = 4)$. 
(b) Same as in panel (a) but for a composite tBLG/hBN system with an additional hBN alignment of $\theta_\mathrm{gr/hBN} = 0.62^\circ$.
For this case there are additional Landau levels emerging from non-integer filling factors of $(\nu_\mathrm{sat} \approx \pm 2.67)$ (see back arrows). 
}
\label{Figure4}
\end{figure}

We simulate the magnetotransport behaviour of the tBLG and the tBLG/hBN system based on the associated band structure. 
To investigate the dependence on magnetic field $B$, we use a Peierl's substitution to include the corresponding gauge phase in each hopping in the tight-binding model.
For the Bloch states we extract the group velocities $v_n (E) = (1/\hbar) \,\partial E_n(k,B)/\partial k$.
We then calculate the magnetoconductance by summing at a fixed energy over all available modes with positive group velocity $v_n(E) > 0$ (i.e., the right-moving modes -- summing over all modes would result in zero due to time reversal symmetry), weighted by the respective group velocity of the underlying Bloch states \cite{Fabian2022Oct} (see \cref{A6} for a derivation)
\begin{equation}
G(E) = \frac{e^2}{h}\frac{\mathrm d}{\mathrm dE}\sum_{\substack{n:\,  v_n(E) > 0\\ E_n < E}} \hbar v_n(E) \Delta k \equiv R^{-1}(E), \label{eq1}
\end{equation}
where $\Delta k$ is the $k$-point spacing of the band structure (we use 5760 $k$-points for the Brillouin zone with a size of 0.25 $\mathrm{nm}^{-1}$).
We compare the magnetoresistance $R(E)$ -- obtained by inverting \cref{eq1} -- of the bare tBLG in \cref{Figure4}(a) to the case of tBLG/hBN in \cref{Figure4}(b) as a function of the filling factor $\nu$ and normalized magnetic flux per tBLG moir\'e unit cell (uc) $\phi = \Phi_\mathrm{uc}/\Phi_0$, where $\Phi_0 = h/e$ is the magnetic flux quantum ($\nu$ and $\phi$ are thus both defined with respect to the tBLG unit cell). 
In both cases, the band gaps caused by the tBLG moir\'e lattice give rise to Landau fans at integer fillings $\nu = \pm 4$, i.e., at full filling of four charge carriers per tBLG moir\'e unit cell.
In the case of tBLG/hBN in \cref{Figure4}(b) the expected additional set of Landau fans emerge from the band gaps induced by the supermoir\'e structure at non-integer filling $\nu \approx \pm 2.67 \equiv \nu_\mathrm{sat}$ (see arrows in \cref{Figure4}).
In our single-particle calculations, this 
fractional filling (of 2 out of three AA sites times 4) emerges because the graphene/hBN moir\'e potential acts as a staggered potential and breaks the degeneracy of the AA sites within the supermoir\'e unit cell.
Crucially, in order to observe an additional band gap at full filling of a selected subset of AA sites, we need the very same partitioning of AA sites in each supermoir\'e unit cell, so we require a \textit{commensurate} moir\' e supercell. \\

\subsection{General selection rule}
To generalize our discussed model we consider a commensurate supermoir\'e structure consisting of $l$ tBLG moir\'e unit cells and $m$ graphene/hBN moir\'e unit cells ($l,m$ both integers), so that the area of the supermoir\'e unit cell is given by $l A_{\mathrm{tBLG}} =  m A_{\mathrm{gr/hBN}}$.
If the graphene/hBN moir\'e lattice selects a subset $p$ of the $l$ tBLG AA sites within the overall supermoir\'e lattice, we expect the appearance of an additional band gap and the accompanying Landau fan at a non-integer filling factor of 
\begin{align}
\nu_{\mathrm{sat}}(l,p) = 4 \frac{p}{l} = 4\frac{p A_{\mathrm{tBLG}}}{m A_\mathrm{gr/hBN}}, \qquad p<l. \label{Eq1}
\end{align}
The geometry of the model system we considered above  defines $l = 3$ and $m = 4$. 
In the corresponding supermoir\'e unit cell, one of the three tBLG AA sites is directly aligned with an AA site of the graphene/hBN moir\'e lattice below, and thus differs from the other two tBLG AA sites (see \cref{A7}). 
This observation suggests that the threefold degenerate AA sites of the unperturbed tBLG are split into a doublet and a singlet.
Consequently, possible values for $p$ are either $p = 1$ or $p = 2$, depending on whether the final energy of the single hBN-aligned tBLG AA site or the two other non hBN-aligned tBLG AA sites is lower. 
The hBN-aligned tBLG AA site is less energetically favorable \cite{PhysRevB.84.195414}, suggesting that the doublet is filled first, and thus $p=2$.
Indeed, the numerical value $\nu_{\mathrm{sat}}\approx 2.67$ is consistent with $p=2$ for our model system, as $\nu_\mathrm{sat}(3,2) = 4 \times 2/3 \approx 2.67$  according to \cref{Eq1}.\\

We find that the appearance of an additional gap is robust with respect to different twist angles of the tBLG moir\'e lattice (see \cref{A6}). 
The geometric constraints of a periodic, commensurate moir\'e lattice allow only for a limited set of combinations of $l$ and $m$ (see \cref{A7}).
In general, we expect values of $l$ and $m$ to be limited by the experimentally realizable size of a regular supermoir\'e lattice, as otherwise twist angle inhomogeneities would wash out any additional Landau fans (see \cref{A4}).
The associated fraction $p/l$ depends on the specific details of the (staggered) potential induced by the graphene/hBN moir\'e lattice within the tBLG, and the relative geometrical orientation of the two moir\'e lattices \cite{Lei2021Jul}:
The energetic shift due to the presence of the graphene/hBN moir\'e lattice originates from its three high-symmetry sites with local energetic shifts~\cite{PhysRevB.84.195414} -- the exact value of $p/l$ will therefore depend on the precise relative alignment of these sites with the AA sites of the tBLG moir\'e lattice.\\

\begin{figure}[!t]
\centering
\includegraphics[draft=false,keepaspectratio=true,clip,width=1\linewidth]{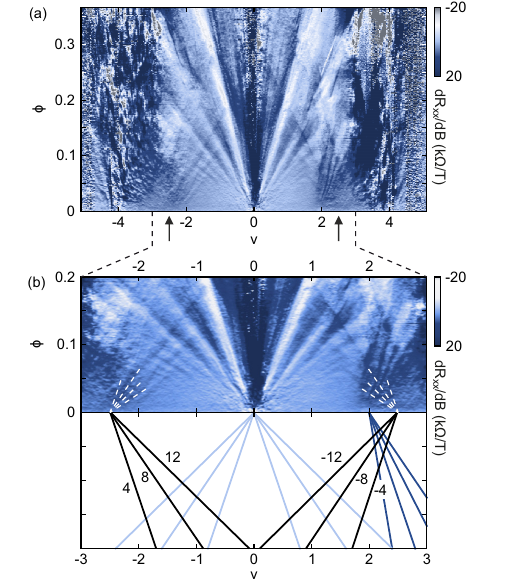}
\caption{ 
(a) Magnetic field derivative of the longitudinal resistance as a function of the filling factor $\nu$ and normalized magnetic flux per tBLG moir\'e superlattice unit cell $\phi$. 
We observe clear Landau levels emerging from charge neutrality and an integer filling factor of $\nu = 2$. 
The Landau levels emerging from $\nu = \pm 4$ are less pronounced.
Furthermore, additional Landau levels emerging from a non-integer filling factor of around $\nu \approx \pm 2.5 \equiv \nu_\mathrm{sat}$ tilted towards the charge neutrality point are visible. 
(b) Zoom-in into the relevant area from panel (a). 
The lines are plotted as a guide to the eye to highlight relevant Landau fans at the charge neutrality point (light blue), the integer filling $\nu = +2$ (dark blue) and at the non-integer filling factor $\nu \approx \pm 2.5$ (black).}
\label{Figure5}
\end{figure}

\section{Experimental Observation}
Next, we turn to transport experiments which show that a Landau fan can indeed be observed at a non-integer filling factor, conclusively due to commensurate tBLG/hBN moir\'e lattices.
In a four-terminal setup, we investigate a tBLG/hBN Hall bar device at a temperature of $T \approx 30$~mK (see \crefrange{A1}{A4} for details on the fabrication and further experimental characterization) and perform magnetotransport measurements which we compare to theory (see \crefrange{Figure4}{Figure5}). 
In Fig.~\ref{Figure5}(a), we show the derivative of the longitudinal resistance with respect to the magnetic field $\mathrm{d}R_\mathrm{xx}/\mathrm{d}B$ as a function of $\nu$ and $\phi$  (see \cref{A4} for a plot without the derivative along the magnetic field axis). 
We observe Landau fans emerging from the charge-neutrality point at $\nu = 0$, and the band insulating states at $\nu = \pm 4$, as well as a set of Landau levels arising from the correlated insulator at $\nu = 2$ -- as it is expected for tBLG near magic angle~\cite{Cao2018Apr, Lu2019Oct}. 
In addition, we also observe a set of Landau fans emerging from non-integer fillings $\nu_\mathrm{sat} = \pm (2.50 \pm 0.11)$ (see Fig.~\ref{Figure5}(b); the fractional fillings where the additional feature occur are marked by the black arrows).
If we attribute the additional Landau fans at filling $\nu_\mathrm{sat}$ to full filling from a hypothetical moir\'e unit cell of size $A_{\mathrm{hyp}}$, we would get $A_{\mathrm{hyp}}  n_\mathrm{sat} = 4$, where $n_\mathrm{sat}$ is the charge carrier density where the Landau fan appears. 
In units of fillings of the tBLG moir\'e unit cell, this results in $A_\mathrm{hyp} = 4A_{\mathrm{tBLG}}/\nu_{\mathrm{sat}} = 1.6 A_{\mathrm{tBLG}}$. 
Assigning this hypothetical moir\'e unit cell $A_{\mathrm{hyp}}$ to a graphene/hBN moir\'e unit cell yields a moir\'e wavelength of $\lambda^\mathrm{gr/hBN}_\mathrm{hyp} \approx  18.5$~nm, which is larger than the theoretical maximal value for a graphene/hBN moir\'e lattice of $\lambda_\mathrm{gr/hBN}^\mathrm{max} \approx 14.7$~nm~\cite{Wang2019Dec}, making this scenario impossible. 
Furthermore, the area $A_{\mathrm{hyp}}$ cannot correspond to any unit cell of the supermoir\'e lattice either, since this would require an area of at least $2 A_{\mathrm{tBLG}}$.
We thus conclude that the observed Landau fans cannot be explained by full filling of a hypothetical cell size.
Instead, the real space selection rule described by~\cref{Eq1} can easily explain the observed additional Landau fans.\\
The question remains whether the observed feature of additional Landau fans at fractional fillings is due to correlations. 
The phase diagram of twisted bilayer graphene is very complex, with a plethora of phenomena driven by many-body effects~\cite{Xie_2021}, some of which could give rise to fractional states. However, here we identify a number of key experimental features that are inconsistent with a many-body explanation:
(i)~The additional Landau fans extend over a wide density range  from $\nu \approx \pm 2.5$ at $\phi=0$ to $\nu \approx \pm 0.5$ at $\phi \approx 0.2$. 
In contrast, reports of correlation-driven Landau fans are extremely limited in carrier  density (as correlated states emerge from the physics of specific fillings). 
The latter is particularly true for the many-body physics of fractional states \cite{Cao2018Apr2, Xie_2021}.  
(ii)~The slope of the Landau fans we observe implies a Landau level degeneracy of $\nu_\mathrm{LL} = -12, -8, -4, 4, 8, 12$, consistent with single-particle predictions but inconsistent with predictions from fractional many-body physics~\cite{Novoselov2006Mar,Engels2014Sep,Schmitz2017Jun, Xie_2021}.
(iii) ~The features of the additional Landau fans emanating from the fractional filling cross through those from the charge neutrality point without interruption, which is not expected for Landau levels emerging from correlated states in tBLG~\cite{Zondiner2020Jun, Yu2022Jul}.
We therefore attribute these additional Landau fans to the presence of the graphene/hBN moir\'e lattice acting as a staggered potential that selects a subset of tBLG AA sites within each supermoir\'e unit cell, as shown by the magnetotransport simulation [see Fig.~\ref{Figure4}(b)].
This also serves as a warning that additional Landau fans do not necessarily imply the emergence of correlated physics.
Indeed, our mean-field tight-binding approach explains the resulting band gap at fractional fillings of the tBLG moir\'e lattice at $\nu_{\mathrm{sat}}$ entirely based on an effective single-body Hamiltonian by the lifting of the degeneracies of the tBLG AA sites, in line with both experimental observations mentioned above.\\

To estimate the tBLG twist angle, $\theta_\mathrm{tBLG}$, we determine the tBLG superlattice density from magnetotransport measurements to be $n_\mathrm{s} \approx \pm (2.17 \pm 0.06) \times 10^{12} \, \mathrm{cm^{-2}}$ and use $\theta_\mathrm{tBLG} = [\sqrt{3}n_\mathrm{s}/8]^{1/2}a$, where $a = 0.246 \, \mathrm{nm}$ is the graphene lattice constant \cite{Cao2018Apr2}.
This results in a value of $\theta_\mathrm{tBLG} \approx 0.97^\circ \pm 0.02^\circ$ (see \cref{A2} for further details).
For the composite graphene/hBN moir\'e lattice, the discussion above suggests a filling of some fraction $p/l$ of the AA sites within the supermoir\'e unit cell.
However, since the filling factor $\nu_\mathrm{sat}$ and the area of the tBLG moir\'e unit cell $A_\mathrm{tBLG}$ are the only experimental accessible quantities in \cref{Eq1}, the size of the graphene/hBN moir\'e unit cell area is not uniquely determined by the measured filling $\nu_\mathrm{sat}$ -- it depends still on the fraction $p/m$ (see \cref{A2}).
Hence, there are triples of integers $(l,m,p)$ which result in identical values for $\nu_\mathrm{sat}$ but correspond to different angles $\theta_\mathrm{gr/hBN}$.
This result might appear surprising, given that the filling at which moir\'e-induced Landau fans appear are usually precise indicators for the underlying moir\'e unit cell areas and thus for the occurring twist angles.
The underlying physical reason is that $\nu_{\mathrm{sat}}$ is not directly determined by the size of the graphene/hBN moir\'e lattice, but rather by the degeneracy lifting induced by the graphene/hBN moir\'e lattice on the AA sites of the tBLG.
Consequently, it highlights the importance of the discussed real-space effect and shows that great care must been taken by the interpretation of moir\'e-induced Landau fans.
In particular, there are values for $\nu_{\mathrm{sat}}$ that are compatible with both, a particular choice of integers $(l,m,p)$ or full filling of a meaningful unit cell area $A_{\mathrm{hyp}}$. 
For example, if $p$ is small, the associated $A_{\mathrm{hyp}}$ becomes sufficiently large to be potentially caused by full filling of the supermoir\'e unit cell.\\

While we cannot uniquely identify the two variables $p$ and $m$ from a measurement of $\nu_{\mathrm{sat}}$, we can identify plausible candidates.
Plausible in this context means that the size of the superstructure is below $50 \, \mathrm{nm} \times 50 \, \mathrm{nm} = 2500 \, \mathrm{nm}^2$, such that the influence of twist angle variations on the satellite features are less likely.
Consequently, we expect small integers for the parameters $l$ and $m$.
Considering simple fractions that correspond to a commensurate moir\'e lattice and yield a $\nu_\mathrm{sat}$ within the experimental error bounds (see \cref{A2}), one finds $(l=13, p = 8)$ yielding according to \cref{Eq1} $\nu_\mathrm{sat} = 4\times 8/13 \approx 2.46$ (see \cref{A7}).
To determine the corresponding value of the parameter $m$, we search for an area ratio yielding a commensurate supermoir\'e lattice, which gives $m= 16$ (see also \cref{FigureS8} in  \cref{A7} for a schematic illustration).
Assuming these values, we can estimate the twist angle of the graphene/hBN moir\'e lattice by using the general geometric relation between the moir\'e unit cell size $\lambda_\mathrm{gr/hBN} = [2A_\mathrm{gr/hBN}/\sqrt{3}]^{1/2} = (13.05 \pm 0.50)$~nm to calculate the twist angle via \cite{Ribeiro-Palau2018Aug}
\begin{align}
    \theta_\mathrm{gr/hBN} = \arccos \left[ 1 - \frac{(1 + \delta)a^2}{2 \lambda_\mathrm{gr/hBN}^2} + \frac{\delta^2}{2(1 + \delta)}\right],
\end{align}
where $\delta \approx 0.017$ is the relative lattice constant mismatch between graphene and hBN.
This results in an extracted twist angle of $\theta_\mathrm{gr/hBN} = 0.50^\circ \pm 0.10^\circ$ ($A_\mathrm{gr/hBN} \approx 147.5 \, \mathrm{nm}^2$).
This value must be treated with caution, as the next largest possible commensurate moir\'e system (just larger by a factor of about 1.3), providing $\nu_\mathrm{sat}$ values within the experimental error bars, is defined by $(l = 13, \, m = 21, \, p=8)$, giving a graphene/hBN twist angle of $\theta_\mathrm{gr/hBN} = 0.79^\circ \pm 0.08^\circ$ ($A_\mathrm{gr/hBN} \approx 112.5 \, \mathrm{nm}^2$).
These different estimates nicely illustrate the difficulty of determining the twist angle in such multi-layer van-der-Waals heterostructures. 
In both cases, the non-integer fraction $\nu_{\mathrm{sat}}$ of full filling of the tBLG unit cell at which we find the additional Landau fans can be expressed as a simple fraction, as predicted by~\cref{Eq1}.\\

Note that the requirement for the appearance of the additional Landau fans resulting from the graphene/hBN moir\'e lattice is commensurability.
In an incommensurate supermoir\'e lattice, one would always select different subsets of tBLG AA sites, which does not result in a well-defined band gap. 
Only in the commensurate case, a sufficient number of repetitions in the resulting superstructure  selects the same, well-defined subsets of tBLG AA sites within each supermoir\'e unit cell, imprinting a band gap on the density of states.
We can investigate the minimal number of repetitions required numerically: we find that a prominent gap in the density of states only emerges upon inclusion of the order of five periodic
cells in one direction.
Estimating the characteristic size of the supermoir\'e unit cell as
$\sqrt{l A_\mathrm{gr/hBN}} = \sqrt{13 \times 147.5}$~nm $\approx 44$~nm results in a minimum length scale of the commensurate area of $\approx 220$~nm, in agreement with the typical size of domains with constant twist angle as reported in Ref.~\cite{Uri2020May}.
We, however, expect that the commensurability also contributes to the stabilisation of the moir\'e geometry suppressing twist-angle inhomogeneities resulting in even significantly larger domains.
This reduces the disorder for the Bloch states in the graphene/hBN moir\'e lattice, making the additional Landau fan more prominent. \\
\section{Conclusion}
In conclusion, our work shows that the alignment of hBN can induce additional periodicity in tBLG-based van-der-Waals heterostructures. 
Our theoretical model shows that the resulting superstructure leads to additional band gaps and Landau fans, which we experimentally observe.
We find that the graphene/hBN moir\'e lattice acts as an additional periodic potential that breaks the symmetry between AA sites of the tBLG moir\'e lattice.
As a consequence, the additional Landau fans do not emerge at full filling of a unit cell of certain size, but instead at full filling of a subset of selected AA sites within the supermoir\'e unit cell.
Thus, our work shows that the effects of moir\'e materials can extend to the length scales of tens of nanometers, and that the appearance of additional Landau fans does not
directly imply an associated supercell of related size. 
In future work, further investigation of such corresponding structures may also shed light on the formation and stabilisation of correlated phases in hBN-aligned tBLG heterostructures.\\

\textbf{Author Contributions}
C.Sc. and F.Li. performed the theoretical simulations. 
A.R. built the device. 
S.T. and F.Le. performed the electron beam lithography. 
A.R. and R.J.D. performed the measurements and analyzed the experimental data. 
A.R., R.J.D., C.Sc., F.Li. and C.St. discussed the experimental and theoretical data with input from B.B. and D.M.K..
K.W. and T.T. supplied the hBN crystals. 
A.R., C.Sc., R.J.D., D.M.K., B.B., F.Li. and C.St. wrote the manuscript. \\

\textbf{Acknowledgements} 
The authors thank F.~Haupt for fruitful discussions. 
This work was supported by the FLAG-ERA grants TATTOOS (437214324) and PhotoTBG (471733165)  by the Deutsche Forschungsgemeinschaft (DFG, German Research Foundation), by the Deutsche Forschungsgemeinschaft (DFG, German Research Foundation) under Germany's Excellence Strategy – Cluster of Excellence Matter and Light for Quantum Computing (ML4Q) EXC 2004/1 – 390534769, within the Priority Program SPP 2244 ``2DMP'' - 443273985 and by
the Helmholtz Nano Facility~\cite{Albrecht2017May}. 
K.W. and T.T. acknowledge support from the JSPS KAKENHI (Grant Numbers 20H00354, 21H05233 and 23H02052) and World Premier International Research Center Initiative (WPI), MEXT, Japan.
C.Sc. acknowledges support as a recipient of a DOC fellowship of the Austrian Academy of Sciences. 
Numerical calculations were in part performed on the Vienna Scientific Cluster VSC4 and VSC5.\\

\textbf{Data availability}
The data supporting the findings of this study are available in a Zenodo repository under \url{https://doi.org/10.5281/zenodo.10847164}.
\newpage
\clearpage
\appendix

\section{Sample fabrication}\label{A1}
\begin{figure}[!thb]
\centering
\includegraphics[draft=false,keepaspectratio=true,clip,width=\linewidth]{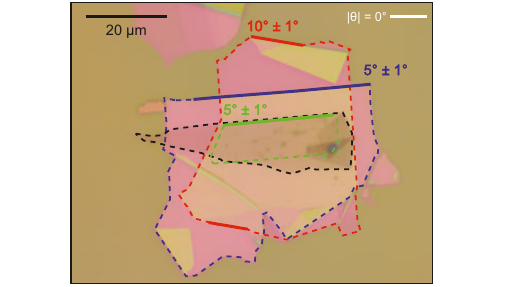}
\caption{Optical image of the fabricated van-der-Waals heterostructure. The dashed lines show the outlines of graphite (black),  tBLG (green), top hBN (red) and bottom hBN (blue) flakes. The graphene/hBN moir\'e lattice is formed on the interface between the bottom hBN and a single-layer graphene flake. Solid lines show the crystallographic axes used for the (mis-)alignment during the fabrication. The angles are given with respect to the horizontal white line. }\label{FigureS1}
\end{figure}
We start our fabrication procedure with exfoliating graphene, graphite (Naturgraphit GmbH) and hBN on standard $\mathrm{Si/SiO_2}$ wafers (oxide thickness: $90$~nm).
Suitable flakes for the further process are identified by optical microscopy.
The fabrication of the tBLG follows the "laser cut and stack" technique \cite{Park2021Apr} using a polycarbonate/polydimethylsiloxane (PC/PDMS) droplet \cite{Kim2016Mar}. 
We adjust a tBLG twist angle of $1.3^\circ$, while the composite hBN/graphene moir\'e lattice is formed by aligning the crystallographic axes of the bottom hBN flake with the axis of the used graphene flake.
We pick-up the hBN flakes (thicknesses of the top and bottom hBN flakes are approximately $25$~nm and $30$~nm, respectively) at $\mathrm{90\, ^\circ C}$, while the graphene pick-ups are performed at $\mathrm{40\, ^\circ C}$. The final pick-up of the graphite back gate is performed at $\mathrm{120\, ^\circ C}$.
The finished stacked is dropped on a marked $\mathrm{Si/SiO_2}$ chip at $\mathrm{165\, ^\circ C}$.
We remove the PC with chloroform and clean the stack with isopropanol.
An optical image of the final stack is shown in \cref{FigureS1}.
The graphite flake is used as a back gate, allowing homogeneous gating and screening of the potential disorder from the substrate \cite{Rhodes2019Jun}.
With standard electron beam lithography and reactive ion etching ($\mathrm{CF_4/O_2}$) techniques we define ohmic contacts to the tBLG \cite{Wang2013Nov, Banszerus2015Jul} which are subsequently evaporated (Cr/Au, 5/50~nm). 
With a further lithography and metal evaporation step, we define the lines to the etched contacts (Cr/Au, 5/50~nm). 
The Hall bar geometry is defined after hard mask evaporation (Al, 60~nm) by a final reactive ion etching step ($\mathrm{SF_6/O_2}$).
Finally, we remove the aluminium hard mask in TMAH (2.38\% in DI water). 

\section{Conversion to density axis, twist angle extractions and error estimations} \label{A2}
To convert the applied back gate voltage to the adjusted charge carrier density,
we extract the lever arm of the graphite back gate $\alpha_\mathrm{bg}$ from the slopes of the visible Landau levels emerging from the charge neutrality point.
The slopes are found by least-squares fitting to the minima in resistance in the magnetotransport data.
The corresponding filling sequence of the Landau levels emerging from the central fan is given by $\nu_\mathrm{LL} = \pm 4, \pm 8, \pm 12$. We calculate the lever arm using:
\begin{align}
B_\mathrm{LL} = \frac{h}{\nu_\mathrm{LL} e} \alpha_\mathrm{bg}V_\mathrm{bg} + \mathrm{const.}
\end{align}
This  results in a numerical value of $\alpha_\mathrm{bg} = (5.148 \pm 0.129) \times 10^{15} \, \mathrm{V^{-1}m^{-2}}$. 
This value is in reasonable agreement with the geometric lever arm expected from a simple plate capacitor model:
\begin{align}
\alpha_\mathrm{bg} = \varepsilon_0 \varepsilon_\mathrm{hBN} \frac{1}{ed},
\end{align}
which yields a value of $\alpha_\mathrm{bg} \approx 6.263 \times 10^{15} \, \mathrm{V^{-1}m^{-2}}$. 
Here, we used $\varepsilon_\mathrm{hBN} = 3.4$ \cite{Pierret2022Jun, Laturia2018Mar} and a thickness of $d \approx 30 \, \mathrm{nm}$ for the bottom hBN flake which was extracted via atomic force microscopy. 
We note, that we observed an extrinsic doping in our device leading to a shift of the charge neutrality point away from $V_\mathrm{bg} = 0$~V. 
During the data analysis we corrected for this extrinsic shift by fitting according to $n = \alpha_\mathrm{bg} (V_\mathrm{bg} - V_\mathrm{bg, off})$ with $V_\mathrm{bg, off} = 112$~mV.
The uncertainty on the experimentally extracted lever arm was used to quantify the uncertainties on the extracted twist angles of the two moir\'e lattices. \\

We estimate the tBLG moir\'e superlattice density to be $n_\mathrm{s} = (2.17 \pm 0.06) \times 10^{12} \, \mathrm{cm}^{-2}$, which gives a moir\'e unit cell area of $A_\mathrm{tBLG} = 4/n_\mathrm{s} = (184.3 \pm 5.1)$~$\mathrm{nm}^2$. 
Via $\theta_\mathrm{tBLG} = [\sqrt{3}n_\mathrm{s}a^2/8]^{1/2}$ this directly propagates to a tBLG twist angle of $\theta_\mathrm{tBLG} = 0.97^\circ \pm 0.02^\circ$. To determine the error, we calculate the twist angle with the density values $n_\mathrm{s}^+ = 2.23 \times 10^{12} \, \mathrm{cm^{-2}}$ and $n_\mathrm{s}^- = 2.11 \times 10^{12} \, \mathrm{cm^{-2}}$ and choose the larger deviation from our mean value as the error bound (this value agrees with the result obtained from Gaussian error propagation).
The additional Landau fan feature emerges at a charge carrier density of $n = (1.36  \pm 0.06) \times 10^{12} \, \mathrm{cm^{-2}} \equiv n_\mathrm{sat} $ which corresponds on the filling factor axis to $\nu_\mathrm{sat} = 4n_\mathrm{sat}/n_\mathrm{s} = 2.50 \pm 0.11$.
Here again, we are estimating the uncertainty by calculating the filling factor via $n_\mathrm{sat}^+$ and $n_\mathrm{sat}^-$ and giving the larger bound (calculation is done with the mean value of $n_\mathrm{s} = 2.17 \times 10^{12} \, \mathrm{cm^{-2}}$).
As stated in the main manuscript, assigning this satellite to full filling of a hypothetical moir\'e unit cell of size $A_\mathrm{hyp}$ would result in $A_\mathrm{hyp} = 4/n_\mathrm{sat} = 4 A_\mathrm{tBLG}/\nu_\mathrm{sat} = (1.60 \pm 0.07)A_\mathrm{tBLG}$.
Associating this hypothetical moir\'e unit cell with a graphene/hBN moir\'e unit cell, results in a moir\'e superlattice wavelength of $\lambda_\mathrm{hyp}^\mathrm{gr/hBN} = [2A_\mathrm{hyp}/\sqrt{3}]^{1/2} \approx 18.5 \, \mathrm{nm} > 14.7 \, \mathrm{nm} \approx \lambda_\mathrm{max}^\mathrm{gr/hBN}$, which is larger than the theoretically maximal value of $\lambda_\mathrm{max}^\mathrm{gr/hBN}$ for a graphene/hBN moir\'e lattice \cite{Wang2019Dec}.
Since $A_\mathrm{hyp} < 2A_\mathrm{tBLG}$ holds, the area $A_\mathrm{hyp}$ can also not correspond to any supermoir\'e lattice. 
We thus conclude that the observed satellite is caused by the discussed real-space effect, yielding to the selection rule
\begin{align}
    \nu_\mathrm{sat}(l,p) = 4 \frac{p}{l} = 4 \frac{p A_\mathrm{tBLG}}{mA_\mathrm{gr/hBN}} \nonumber,\\ \qquad lA_\mathrm{tBLG} = mA_\mathrm{gr/hBN}, \qquad p<l, \label{b3}
\end{align}
were $l$ and $m$ are the numbers of the tBLG and graphene/hBN unit cells and $p$ is the number of selected tBLG AA sites.
The experimental observed value of $\nu_\mathrm{sat}$ suggests a priori simple fractions like $p/l = 5/8, 7/11$, or $8/13$ which are all within the experimental uncertainty.
In the next step, we consider the possible ratios $m/l$ of the commensurate areas of the two moir\'e unit cells. 
The simulated ratio of $A_\mathrm{tBLG}/A_\mathrm{gr/hBN} = m/l = 4/3$ implies for both possible values of $p = 1,2$ either $\nu_\mathrm{sat}(l,p) = \nu_\mathrm{sat}(3,2) \approx 2.67$, which is close to the estimated error bound or $\nu_\mathrm{sat}(l,p) = \nu_\mathrm{sat}(3,1) \approx 1.33$ which is far off. 
There are, however, other commensurate unit cell structures with rations $A_\mathrm{tBLG}/A_\mathrm{gr/hBN} = 16/13$ or $21/13$ resulting in $\nu_\mathrm{sat}(l,p) = \nu_\mathrm{sat}(13,8) \approx 2.46$, which are both in line with the experimental value. 
In principle, the experimental value of $\nu_\mathrm{sat} \approx 2.50$ implies for the ratio $m/l$ a denominator $l$ of, e.g. 8 ($4p/l = 4 \times 5/8 = 2.5$), 11 ($4 \times 7/11 = 2.54$) or 13 ($4 \times 8/13 = 2.46$). 
We note, that for $l = 8, 11$, there are no commensurate moir\'e superstructures with a reasonable graphene/hBN moir\'e unit cell size (see~\cref{A7}).
This leaves a fraction of $A_\mathrm{tBLG}/A_\mathrm{gr/hBN} = m/l = 16/13$ for the area ratio as the closest fit to the experimental result for $\nu_{\mathrm{sat}}$.
A further culprit for the remaining differences between theory and experiment might also be limitations of our model geometry, which does not allow for additional corrugations of the tBLG due to the additional hBN/graphene moir\'e lattice, for shearing of the moir\'e unit cell that affects the symmetry of the reconstruction or relaxation effects in the tBLG/hBN heterostructure. 
Consequently, the simplest case $m/l = 4/3$, slightly outside of the experimental error bars, might potentially also be relevant.\\

The twist angle of the graphene/hBN moir\'e lattice can now be estimated for different possible cases. 
To do this, we trace back the graphene/hBN unit cell area to the tBLG moir\'e unit cell area via $A_\mathrm{gr/hBN} = 4pA_\mathrm{tBLG}/(m \nu_\mathrm{sat})$. 
For the two cases ($l = 3, m = 4, p = 2$) and ($l = 13, m = 16, p = 8$) this leads to identical results (since $p/m = 2/4 = 8/16 $) with a numerical value of $A_\mathrm{gr/hBN} = (147.5 \pm 11.1)$~$\mathrm{nm}^2$ (we estimate the uncertainty by using $A_\mathrm{tBLG}^+ = 189.4 \, \mathrm{cm^{-2}}$ and $\nu_\mathrm{gr/hBN}^- = 2.39$ to calculate $A_\mathrm{gr/hBN}^+ = 158.6 \, \mathrm{nm^2}$ and vice versa for $A_\mathrm{gr/hBN}^- = 137.4 \, \mathrm{nm^2}$; we then choose the larger deviation from the mean value to define our error bound). 
This results is equivalent to an area ratio of $A_\mathrm{tBLG}/A_\mathrm{gr/hBN} = 2\nu_\mathrm{sat}/4 = 1.25 \pm 0.06$ (with $\nu_\mathrm{sat} = 2.50 \pm 0.11$).
Now, by exploiting $\lambda_\mathrm{gr/hBN} = [2A_\mathrm{gr/hBN}/\sqrt{3}]^{1/2}$ we calculate (error estimation again via $A_\mathrm{gr/hBN}^\pm$) a numerical value of $\lambda_\mathrm{gr/hBN} = (13.05 \pm 0.50) \, \mathrm{nm}$, which results finally via~\cite{Ribeiro-Palau2018Aug}
\begin{align}
    \theta_\mathrm{gr/hBN} = \arccos \left[1 - \frac{(1 + \delta)a^2}{2 \lambda_\mathrm{gr/hBN}^2} + \frac{\delta^2}{2(1 + \delta)}\right]
\end{align} 
in a graphene/hBN twist angle of $\theta_\mathrm{gr/hBN} = 0.50^\circ \pm 0.10^\circ$ (again estimated by performing the calculation with $\lambda_\mathrm{gr/hBN}^\pm$).
The third case discussed in the main manuscript ($l = 13, m = 21, p = 8$) yields in analogous calculation for the graphene/hBN moir\'e unit cell area a value of $A_\mathrm{gr/hBN} = (112.4 \pm 8.5)$~$\mathrm{nm}^2$ (area ratio $A_\mathrm{tBLG}/A_\mathrm{gr/hBN} = 1.64 \pm 0.07$), corresponding to a moir\'e superlattice wavelength of $\lambda_\mathrm{gr/hBN} = (11.39 \pm 0.44)$~nm. This results in a corresponding twist angle of $\theta_\mathrm{gr/hBN} = 0.79^\circ \pm 0.08^\circ$. 
Further possible commensurate cases up to ($l = 13, m = 25$) are listed in Tab.~\ref{STabular1} in~\cref{A7}.\\

\section{Initial characterization of the device} \label{A3}
\begin{figure*}[!ht]
\centering
\includegraphics[draft=false,keepaspectratio=true,clip,width=1\linewidth]{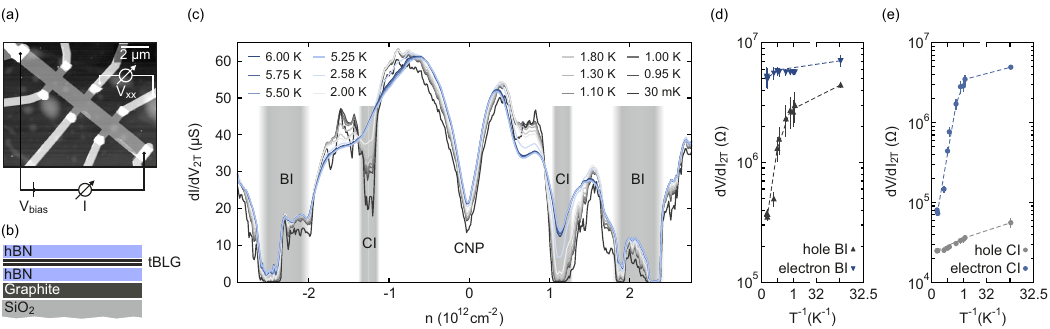}
\caption[Fig2_AR_v4]{(a) Atomic force microscopy image of the measured device. The electrical connections for the two- and four-terminal measurements are depicted. (b) Schematic cross-section of the device showing the composition of the heterostructure. (c) Two-terminal differential conductance $\mathrm dI/\mathrm dV_\mathrm{2T}$ as a function of the charge carrier density $n$ for different temperatures measured along the entire Hall bar structure. The minima in conductance correspond to the band insulators (BI), correlated insulators (CI) and the charge neutrality point (CNP) (d) Arrhenius plot of the differential resistance $\mathrm dV/\mathrm dI_\mathrm{2T}$ of the band insulating states as a function of the inverse temperature. The individual datapoints are extracted by taking the mean of the minima in the band insulating states. The errorbars represent the standard deviations. (e) Same as in panel (d), but for the correlated insulating states.}  
\label{FigureS2}
\end{figure*}
The experiments are performed in a ${}^3\mathrm{He}/{}^4\mathrm{He}$ dilution refrigerator at a base temperature of around $30 \, \mathrm{mK}$ using standard DC and low-frequency lock-in measurement techniques. 
We initially characterize our device by temperature-dependent transport measurements. 
An atomic force microscopy image of the sample is shown in Fig.~\ref{FigureS2}(a), while Fig~\ref{FigureS2}(b) shows a cross-section of the van-der-Waals heterostructure.  
To verify the existence of a tBLG moir\'e lattice, we start characterizing our device by measuring the two-terminal differential conductance $\mathrm{d}I/\mathrm{d}V_\mathrm{2T}$ as a function of the charge carrier density $n$ adjusted by the graphite back gate voltage by applying a symmetric AC bias of $V_\mathrm{ac} = 100\, \mu\mathrm{V}$ along the entire Hall bar structure [see Fig.~\ref{FigureS2}(a)] and separately measuring the current via an in-house built $IV$-converter (gain: $10^7 \, \mathrm{VA^{-1}}$) for different temperatures $T$ [see Fig.~\ref{FigureS2}(c)].
With decreasing temperature we observe multiple dips in differential conductance at certain charge carrier densities, which can be assigned to both, correlated insulating states (CI) and band insulating states (BI) \cite{Cao2018Apr, Cao2018Apr2, Lu2019Oct}. 
From the rapid decrease of differential conductance as a function of temperature, we identify the insulating features around 
$n_\mathrm{s}/2 \approx \pm 1.085 \times 10^{12} 
\, \mathrm{cm^{-2}}$ as correlated insulating states [see also the Arrhenius plot in Fig.~\ref{FigureS2}(e)]. Here, $n_\mathrm{s}$ denotes the superlattice density, i.e. the density which we associate with the edges of the flat bands.
Consequently, the dips in differential conductance around $n_\mathrm{s} \approx \pm 2.17 \times 10^{12} \, \mathrm{cm^{-2}}$ correspond to the band insulating states (BI) or full filling of four holes/electrons of the tBLG moir\'e superlattice unit cell while the correlated insulating states at $n \approx \pm 1.085 \times 10^{12} 
\, \mathrm{cm^{-2}}$ correspond to half-filling of two holes/electrons of the tBLG moir\'e superlattice unit cell [see Fig.~\ref{FigureS2}(c) and the Arrhenius plot in Fig.~\ref{FigureS2}(d)].
The hole doped band insulator shows a plateau in differential conductance emerging at a charge carrier density of $n\approx -2 \times 10^{12} \, \mathrm{cm^{-2}}$ until the minimum is reached at $n  \approx -2.4 \times 10^{12} \, \mathrm{cm^{-2}}$. 
Likewise, the band insulator at electron doping appears to be split in two separate insulating features around charge carrier densities of $n  \approx 1.95 \times 10^{12} \, \mathrm{cm^{-2}}$ and $n \approx 2.3 \times 10^{12} \, \mathrm{cm^{-2}}$.
\begin{figure}[!h]
\centering
\includegraphics[draft=false,keepaspectratio=true,clip,width=\linewidth]{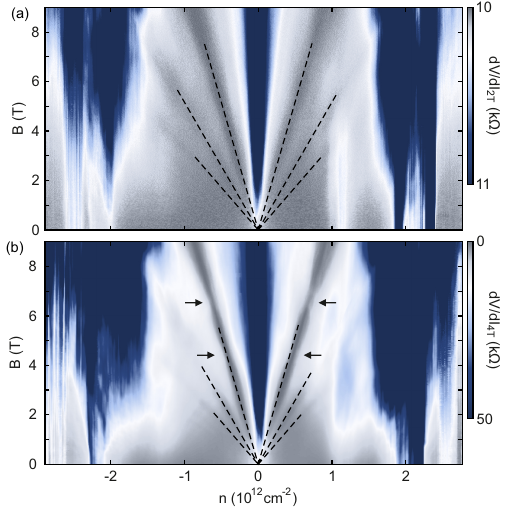}
\caption{(a) Two-terminal differential resistance as a function of magnetic field and carrier density measured along the entire Hall bar structure (compare with \cref{FigureS2}(a). (b) Four-terminal differential resistance measured in the configuration shown in \cref{FigureS2}(a) as a function of magnetic field and carrier density.
The black arrows mark the positions where the Landau levels emerging from the graphene/hBN moir\'e lattice intersect with the Landau levels emerging from the tBLG moir\'e lattice.
Visible is the narrowing of the lowest Landau level emerging from the charge neutrality point at the certain intersection positions.}\label{FigureS3}
\end{figure}
\begin{figure*}[!t]
\centering
\includegraphics[draft=false,keepaspectratio=true,clip,width=\linewidth]{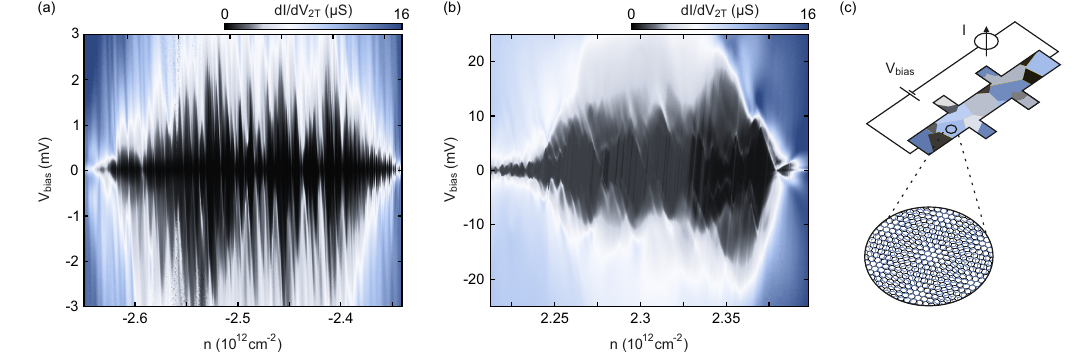}
\caption{ (a) Finite bias spectroscopy on the hole-doped BI state showing individual Coulomb diamonds. (b) Same as in (a) but for the electron-doped BI. (c) Schematic illustration of the measurement setup.}\label{FigureS4}
\end{figure*}
We attribute this distorted shape to twist-angle variations along our device \cite{AUri2020May, Schapers2022Jul, Lau2022Feb}, preventing a clean gap opening \cite{Icking2022Nov}. 
This interpretation is supported by the comparison of two- and four-terminal magnetotransport data in \cref{FigureS3} where the splitting of the electron-doped band insulator is only visible for the two-terminal measurement and additional finite bias spectroscopy measurements on the band insulating states [see \cref{FigureS4}(a,b)].
The temperature activated transport data of the band insulating gaps [Fig.~\ref{FigureS2}(d)] shows a qualitative asymmetry between the hole and electron doped site. 
Over the accessible temperature range, the hole doped band insulator shows a clear thermal activation while the corresponding band insulator at electron doping is barely affected up to the maximum temperature of $T = 6 \, \mathrm{K}$.
This asymmetry is indicative of a difference in the size of the energy gap between electron and hole side, in accordance with the band structure calculations shown in \cref{Figure2} (see also \cref{FigureS7}).
\section{Influence of twist-angle disorder}\label{A4}

Twist angle variations along the device are visible in magnetic field- and bias-dependent measurements.
In Fig.~\ref{FigureS3} we compare the two-terminal differential resistance measured as a function of magnetic field and carrier density along the entire Hall bar structure [panel (a)] and the corresponding four-terminal measurement of the longitudinal resistance [panel (b)].
The splitting of the electron-doped band insulator around $n \approx 2 \times 10^{12} \, \mathrm{cm^{-2}}$ is absent in the four-terminal measurement indicating that it is caused by twist-angle variations along the device.
Furthermore, in the four-terminal data, narrowing features are present in the most prominent Landau level emerging from the charge neutrality point (see black arrows). 
These features correspond to the intersection points between the fan emerging from charge neutrality and the additional hBN-induced Landau fan. 
Therefore, they might indicate the presence of Bloch states (Brown-Zak fermions) caused by restoring the translational symmetry in the graphene/hBN moir\'e lattice. 
Note, that the indications of a graphene/hBN moir\'e lattice are only visible in the four-terminal data which might be explained with a rather homogeneous twist angle distribution in that particular area of the device.
In \cref{FigureS4}(a,b) we show two-terminal finite bias spectroscopy measurements of the band insulating (BI) states.
We do not observe a single diamond-shaped feature of suppressed differential conductance but instead the formation of a multitude of individual Coulomb diamonds which are most prominent in the hole regime. 
We attribute this to the formation of different moir\'e domains in the tBLG superlattice with slightly different twist-angles as sketched in \cref{FigureS4}(c).
The electron-doped BI seems to be slightly less affected from the twist-angle disorder. 
This might be due to stronger stability of this gap compared to the hole-doped BI, which was also observed in the temperature dependence. 
This is noticeable in higher bias values which are necessary to break through the gap.
The difference in the bias values necessary to break through the gaps for both dopings also indicates the mentioned asymmetry visible in the band structure calculation.

\section{Modeling two moir\'e lattices in unison} \label{A5}
\begin{figure}[!ht]
\centering
\includegraphics[draft=false,keepaspectratio=true,clip,width=0.9\linewidth]{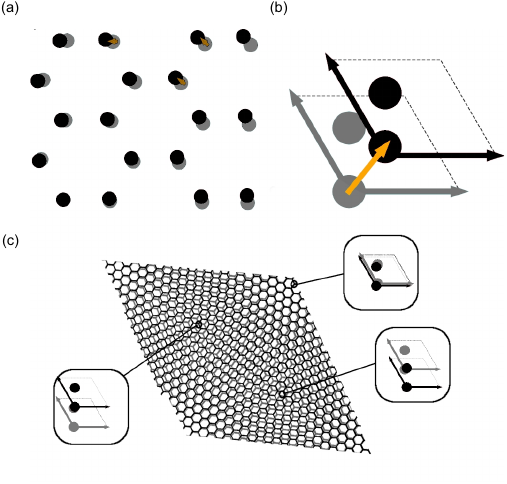}
\caption{ 
(a) Close-up of twisted bilayer graphene (large angle for clarity (black: top layer, gray: bottom layer). (b) Unit cell in the $x$-$y$ plane with local displacement vector $\vec{d}$. 
(c) Unit cell of tBLG.
Each local configuration can be mapped to a unit cell (see insets) with a corresponding shift.
}
\label{FigureS5}
\end{figure}
\begin{figure}[tbh!]
\centering
\includegraphics[draft=false,keepaspectratio=true,clip,width=\linewidth]{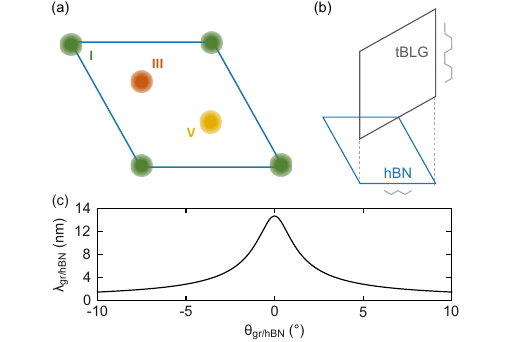}
\caption{ 
(a) Schematic of the real space moir\'e superlattice unit cell of the graphene/hBN system. Colored circles (I,II,V) correspond to centers of Gaussians used in the effective moir\'e potential. 
(b) Schematic explanation of the superposition of the two moir\'e lattices. Light grey lines indicate edge character of the unit cells. Assuming a small graphene/hBN twist angle of $\theta_\mathrm{gr/hBN} \approx 0.62^\circ$ brings the two moire cells to the same periodicity.
(c) Lattice constant of the graphene/hBN moir\'e lattice as a function of their relative twist angle $\theta_{\text{gr/hBN}}$.
}
\label{FigureS6}
\end{figure}
We combine an atomistic tight-binding Hamiltonian parametrized from DFT calculations with a continuum elasticity model that accounts for lattice relaxation in tBLG with an effective symmetry breaking potential that accurately captures the influence of alignment to an hBN layer.
The local stacking configuration of the two rotated graphene layers (parametrized via a displacement vector $\vec{d}$) changes throughout the large moir\'e super cell (\cref{FigureS5}).
We sample the two-dimensional configuration space of the displacement vector $\vec{d}$ with a $10\times 10$ grid of DFT calculations for periodic, primitive-cell calculations mapping the configuration space of different displacements $\vec{d}$ [\cref{FigureS5}(b)].
We obtain tight binding parameters for intermediate stackings using Fourier interpolation.
We can then map the different local stackings within the large supercell to the corresponding tight-binding parameters.
For the primitive bilayer DFT calculations, we use VASP \cite{vasp1} with LDA, $25 \times 25$ Monkhorst $\vec{k}$-space grid and a plane wave cutoff of $380$~eV. 
Atomic positions are relaxed in out-of-plane direction but fixed in-plane.
We then project all Kohn-Sham orbitals onto one $p_{\text{z}}$ orbital per carbon site (via Wannier90 \cite{wann1,wann2,wann3}) and thus capture all the influence of the tBLG moir\'e lattice that is relevant for electronic transport close to the Fermi energy.
From this point on we can smoothly interpolate tight-binding couplings $\gamma_{ij}$ across the entire moir\'e supercell. 
However, we have yet to account for the significant strain fields of the real moir\'e supercell. 
We do so via exponential correction factors that depend on the local strain:
\begin{equation}
\gamma_{i,j}^{\text{(corr.)}}  = \gamma_{i,j} \text{e}^{-\Delta l_{ij} \alpha_{ij}},
\end{equation}
where $\Delta l_{ij}$ is the change in inter orbital distance due to mechanical relaxation and $\alpha_{ij}$ encodes the distance sensitivity of individual tight-binding hopping parameters. 
We determine the $\alpha_{ij}$ from a set of DFT calculations on primitive unit cells of strained single layer graphene with subsequent Wannierization. 
Finally, we determine the $\Delta l_{ij}$ via an approach that resembles the elasticity models of Nam and Koshino \cite{Nam2017Aug}. 
Such an approach determines an equilibrium configuration that balances energy gain due to more favourable stacking fault energies with the elastic energy cost associated with in-plane displacements. 
The corresponding energy functional is of the form,
\begin{equation}
U_{\text{tot}} = U_{\text{E}}[\vec{u}_{\text{bot}}] + U_{\text{E}}[\vec{u}_{\text{top}}] + U_{\text{B}}[\vec{u}_{\text{bot}}, \vec{u}_{\text{top}}],
\label{e:manch1}
\end{equation}

\noindent where $\vec{u}_{\text{bot/top}}$ are the local displacement vectors in bottom and top layer respectively. 
The third, stacking dependent term is elegantly expressed via the first few Fourier components $c_{\vec{G}}$ of the generalized stacking fault energy:
\begin{equation}
U_{\text{B}}[\vec{u}_{\text{bot}}, \vec{u}_{\text{top}}] = \int \sum_{\vec{G}} c_{\vec{G}} \text{e}^{\text{i} \kl{(\vec{d}+\vec{u}_{\text{bot}}-\vec{u}_{\text{top}}) \cdot \vec{G}}} \text{d}\vec{r}
\end{equation}
with $\vec{G}$ running over reciprocal lattice vectors. 
The intralayer contributions $U_{\text{E}}$ of both layers (m$=$ top/bot) read, 
\begin{widetext}
\begin{align}
U_{\text{E}}[\vec{u}^{\text{m}}] = \int 
\Biggl[ &\frac{\lambda+\mu}{2} \kl{ \frac{\partial u_x^{\text{m}} }{\partial x}+\frac{\partial u_y^{\text{m}} }{\partial y} }^2+\frac{\mu }{2} \kl{ \kl{ \frac{\partial u_x^{\text{m}} }{\partial x} \shortminus \frac{\partial u_y^{\text{m}} }{\partial y} }^2 +
\kl{ \frac{\partial u_x^{\text{m}} }{\partial y}+\frac{\partial u_y^{\text{m}} }{\partial x} }^2} \Biggl]\text{d}\vec{r}
\end{align}
\end{widetext}
with Lam\'e parameters $\lambda = 3.25 \, \mathrm{eV\AA^{\shortminus2}}$ and $\mu = 9.57 \, \mathrm{eV\AA^{\shortminus2}}$.
We then solve the Euler-Lagrange equations of the system following closely along the procedures in \cite{Nam2017Aug}. 
The main result of this relaxation is the proliferation of the energetically favourable AB stacking region in the upper left half of the moir\'e supercell.

\begin{figure*}[!th]
\centering
\includegraphics[draft=false,keepaspectratio=true,clip,width=1\linewidth]{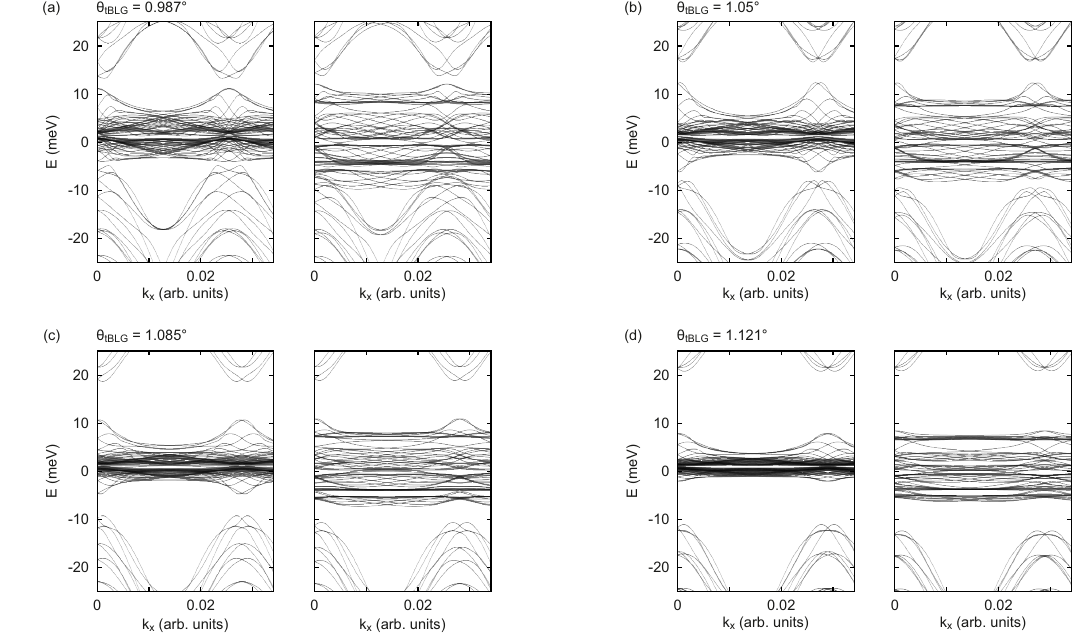}
\caption{Band structure calculations for different twist angles without (left subpanels) and with (right subpanels) hBN alignment.} 
\label{FigureS7}
\end{figure*}

We next consider models for the additional second moir\'e lattice between hBN and one graphene sheet of the tBLG.
Due to the wide band gap of hBN, there is no need to explicitly model the nitrogen and boron atoms themselves. 
Instead, we use additional on-site corrections of the $p_z$ orbitals of the graphene sheet. 
We describe the effect of the hBN via a slowly varying background potential $V_i$ and a short-range symmetry breaking potential $W_i$ \cite{PhysRevB.90.165404}: 
\begin{align}
		V_{\text{hBN}} &= \sum_{i = \text{I,III,V}} V_i \text{e}^{\shortminus\frac{\kl{\mathbf{r}\shortminus\mathbf{R}_i}^2}{2w_i^2}} \cdot \sigma_0 \otimes \tau_0 \nonumber\\&+ \sum_{i = \text{I,III,V}} W_i \text{e}^{\shortminus\frac{\kl{\mathbf{r}\shortminus\mathbf{R}_i}^2}{2w_i^2}} \cdot \sigma_z \otimes \tau_0,
	\label{e:laris}
\end{align}
where $w_i$ are characteristic length scales, $\sigma_0(\sigma_z)$ represent unity ($z$-Pauli matrix) in sublattice space and $\tau_0$ is the unity in valley space. 
The $W_i$ terms proportional to $\sigma_z$ reflect the short-range variations in local chemical environment between adjacent carbon atoms (e.g., in a configuration where one carbon atom sits on top of a nitrogen and the other on a boron atom.
While the $V_i$ break particle-hole symmetry the $W_i$ introduce sublattice asymmetry.
Derived from DFT calculations \cite{PhysRevB.89.161401,PhysRevB.84.195414} these terms can accurately describe the effect of hBN alignment with respect to one of the graphene layers. 
We partition the graphene/hBN moir\'e unit cell in five regions based on relative local alignment with the graphene layer (see \cref{FigureS6}).
Assigning different amplitudes ($V_{\text{I}} =V_{\text{III}} =0$~meV, $V_{\text{V}} = 100$~meV, $W_{\text{I}} = 57$~meV, $W_{\text{III}} = \shortminus 34 $~meV, $W_{\text{V}} =\shortminus 47$~meV) and widths ($0.63w_{\text{I}}=w_{\text{III}}=w_{\text{V}}=7$~nm) to the Gaussians in \cref{e:laris} allows us to effectively model the influence of the hBN alignment and introduces the second moir\'e lattice length scale into the tBLG Hamiltonian in an elegant and easily adaptable manner. 
This adaptability is very important for us to create structures with feasible periodicity. 
The relaxed displacement-mapping method we use for the derivation of the tBLG tight-binding Hamiltonian can only be applied to commensurate twist angles $\theta_{\mathrm{tBLG}}$.
Having access to a set of different twist angles of tBLG Hamiltonians we identify $\theta_\mathrm{tBLG} = 0.987^{\circ}$ as the closest one to the experimentally determined twist angles. 
The unit cell for this moir\'e system is spanned by the vectors $\mathbf{a}_1=(14.3, 0)^T$~nm and $\mathbf{a}_2=(\shortminus7.1 , 12.4)^T$~nm and features armchair borders.
The effective hBN potential is derived for a perfectly aligned graphene/hBN moir\'e cell ($\mathbf{b}_1=(13.8 , 0)^T$~nm, $\mathbf{b}_2=(\shortminus6.9 , 11.9)^T$~nm) that features zig-zag borders. 
Their difference in border character is easily reconciled via a rotation.
However, this would result in a slightly different periodicity in $x$ direction.
We avoid cumbersome duplication to their least common multiple and instead assume a slight rotation of the graphene/hBN moir\'e lattice. 
This bilayer of materials with unequal lattice constants also features an angle dependence \cite{PhysRevB.90.155406} (see \cref{FigureS6}c):
\begin{equation}
	\lambda_\mathrm{gr/hBN} = \frac{(1+\delta)a}{\sqrt{\delta^2+2\kl{1+\delta}\kl{1-\cos\theta_\mathrm{gr/hBN}}}}
\end{equation}
where $\delta$ is the relative mismatch of lattice constants. 
A small twist angle $\theta_\mathrm{gr/hBN} = 0.62^{\circ}$ results in perfect agreement of $x$ periodicity for the composite moir\'e system which we can then duplicate in $y$ direction to describe ribbons of realistic width.
Terminating the edges with $\sigma_z$ potentials in sublattice space suppresses surface states.
\section{Band structure and transport calculations} \label{A6}
We efficiently calculate the band structure of a ribbon of several of the --- by themselves already quite sizable --- moir\'e unit cells with methods we developed in the study of graphene/hBN moir\'e lattices \cite{Fabian2022Oct}.
We partition the final entire tight-binding Hamiltonian into the (small) part defining the periodic boundary conditions in $x$-direction, $H_I$ and the rest, $H_0$.
Bloch states of our structure thus follow
\begin{equation}\label{eBandstruct}
\left[H_0 + e^{\mathrm ik\Delta x}H_I + e^{-\mathrm ik\Delta x}H_I^\dagger\right] \psi_n  = E_n \psi_n.
\end{equation}
To solve this equation, we employ iterative methods for the eigenvalues close to charge neutrality. 
We solve via shift-and-invert in combination with the Lanczos method \cite{Lanczos1950}. 
In order to partially avoid cubic scaling we perform several independent matrix factorizations around different energies to eventually cover the range $E \in [-0.25\, \mathrm{eV}, 0.25\, \mathrm{eV}]$.
The accuracy of subsequent evaluations of the conductivity depends on the sampling resolution in reciprocal space (no further improvements noticeable beyond $N_{\text{kpt}} > 3000$ in our system).
We optimize sampling at such high $\mathbf{k}$-point densities by exploiting the continuity of bands along small distances in $\mathbf{k}$-space. 
To this extent we avoid solving the Bloch eigenvalue problem at most of the $N_{\text{kpt}}$ $\mathbf{k}$-points and instead span a Krylov space for a subset $N_{\text{pillars}}$ of ``pillar'' $\mathbf{k}$-points.
Combining the Krylov spaces of two adjacent pillar points $k_j$ and $k_{j+1}$ generates a basis $\{b_i\}$ on which to project for all intermediate $\mathbf{k}$-points $k'$ ($k_j < k' < k_{j+1}$) and thus evaluate band energies.
Main caveat of this approach is the emergence of unphysical eigenvalues due to the artificially enlarged size of the combined Krylov spaces.
We remedy this by evaluating an error norm which sees the solutions projected onto a fixed set of randomly chosen vectors $\phi_i \in \mathbb{C}^N$. 
This error measure vanishes for an eigenstate of the full problem and thus allows filtering out unphysical solutions.
Matrix vector operations between the $\phi_i$ and $H_0$ or $H_I$ do not depend on $\mathbf{k}$ and thus need only be evaluated once.
In Fig.~\ref{FigureS7} we show the evolution of the band structure for the non-aligned (left subpanels) and aligned (right subpanels) cases with increasing tBLG twist angle.\\

To express the group velocity $v_n$, we derive \cref{eBandstruct} with respect to $k$ and form an expectation value,
\begin{equation}\label{evg}
v_n = \frac{1}{\hbar}\frac{\partial E}{\partial k} = \frac{\mathrm i\Delta x}{\hbar}\phi_n^\dagger \left[e^{\mathrm i k \Delta x} H_I - e^{-\mathrm i k \Delta x} H_I^\dagger \right] \phi_n.
\end{equation}  
To calculate the magnetoconductance, we include a magnetic field via Peierl's substitution. 
For an infinite ribbon, the conductance is simply given by the number of open modes $M_+(E)$ in positive $x$-direction times the conductance of each mode, $e^2/h$,
\begin{equation}\label{eq:GE}
G(E) = \frac{e^2}{h} M_+(E).
\end{equation}
To approximate $M_+(E)$, we consider the integrated density of states of Bloch eigenstates moving in a single direction in a certain energy interval $\Delta E$,
\begin{equation}
 \rho_+(E)  = \sum_{\substack{n: \, v_n(E) > 0\\ E_n < E}} 1.
\end{equation}
 $\rho_+(E)$ is related to $M_+(E)$ via
\begin{equation*}
 M_+(E) = \frac{\rho_+(E+\Delta E) - \rho_+(E)}{\Delta E} \Delta E\approx \frac{\partial \rho_+(E)}{\partial E} \Delta E
\end{equation*}
for a small interval $\Delta E$.
When numerically evaluating Bloch eigenstates, one typically evaluates eigenenergies by solving Bloch’s equation \eqref{eBandstruct} on an equidistant $k$-grid.
It is therefore convenient to replace
\begin{equation*}
\Delta E \approx \frac{\partial E_n}{\partial k}\Delta k = \hbar v_n \Delta k
\end{equation*}
Inserting these equations into \eqref{eq:GE} yields an estimate for the conductance $G(E)$ \cite{Fabian2022Oct},
\begin{align}
	    G(E) \approx \frac{e^2}{h} \frac{\mathrm d}{ \mathrm dE} \sum_{\substack{n: \, v_n(E) > 0\\ E_n < E}} \hbar v_n(E) \, \Delta k.
\label{eTransportfromBand}
\end{align}

The use of Eq.~(\ref{eTransportfromBand}) greatly reduces the computational cost compared to calculation of $G(E)$ using the Landau-B\"uttiker formalism while yielding nearly identical results~\cite{Fabian2022Oct}. 
As our simulation contains the full information on the band structure, the energy axis can be readily transformed into the charge carrier density simply by integrating the density of states, 
\begin{equation*}
n(E) = \int_0^{E} \rho(E^\prime) \, \mathrm dE^\prime,
\end{equation*}
obtained from counting the bands with appropriate weight based on their group velocity. 
We set $n=0$ at the charge neutrality point $E=0$ of graphene and express $n$ in units of $n_0 = 1/S$, the density of one electron per tBLG moir\'e unit cell. 
Likewise, the magnetic field strength $B$ is conveniently expressed in units of magnetic flux quanta through one moir\'e supercell $\Phi/\Phi_0$.
        
\section{Commensurate moir\'e structures} \label{A7}
\begin{figure*}[!t]
\centering
\includegraphics[width=0.95\linewidth]{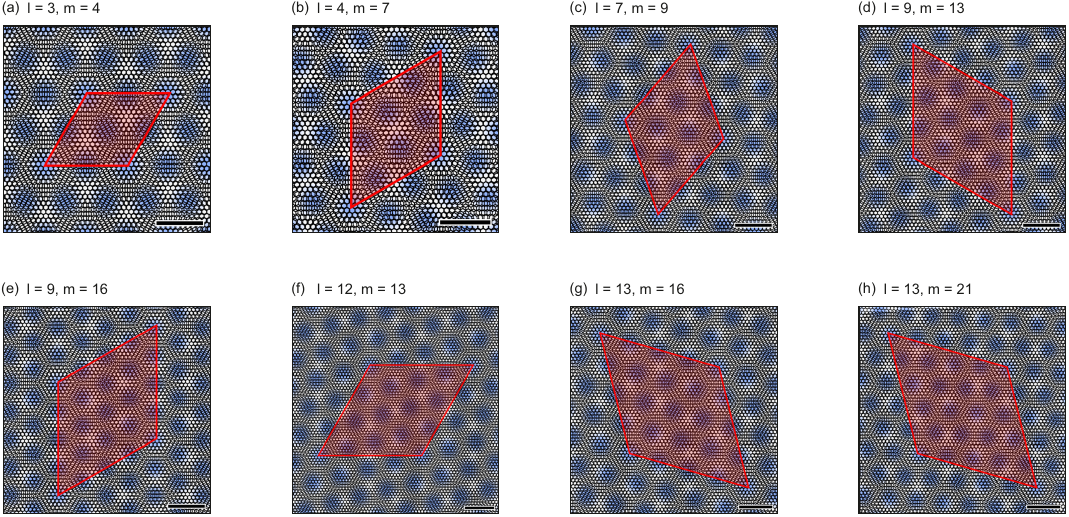}
\caption{Selection of commensurate moir\'e structures listed in Table~\ref{STabular1}.
Shown is the tBLG moir\'e lattice (black) as well as the AA site of the graphene/hBN moir\'e lattice (blue dots). 
Depending on the choice of $l$ and $m$ we receive a commensurate structure for different orientations $\varphi$ of the two moir\'e lattices with respect to each other. 
The unit cell of the resulting supermoir\'e lattice is depicted as a red rhombus. 
The black scale bar corresponds to the spacing of the tBLG AA sites.} 
\label{FigureS8}
\end{figure*}

\begin{table}[!th]
\centering
\begin{tabular}{ccc|c|c|c|c}
$l$ & $m$ &$p$ & $A_\mathrm{sm} \, \mathrm{[nm^2]}$ & $\nu_\mathrm{gr/hBN} = 4 p/l$ & $\theta_\mathrm{gr/hBN} \, \mathrm{[{}^\circ]}$ & $\varphi$ [${}^\circ$] \\
\hline
3 & 4 & 2 & 530 & 2.67 & $0.50 \pm 0.10$ & 90\\
4 & 7 & 3 & 706 & 3 & $0.67 \pm 0.08$ & 100\\
7 & 9 & 4 & 1236 & 2.28 & $0.63 \pm 0.09$ & 100\\
9 & 13 & 6 & 1589 & 2.67 & $0.59 \pm 0.09$ & 73\\
9 & 16 & 6 & 1589 & 2.67 & $0.81 \pm 0.08$ & 60\\
12 & 13 & 8 & 2119 & 2.67 & $0.18 \pm 0.15$ & 76\\
12 & 19 & 8 & 2119 & 2.67 & $0.69 \pm 0.08$ & 76\\
13 & 16 & 8 & 2296 & 2.46 & $0.50 \pm 0.10$ & 106\\
13 & 19 & 8 & 2296 & 2.46 & $0.69 \pm 0.08$ & 82\\
13 & 21 & 8 & 2296 & 2.46 & $0.79 \pm 0.08$ & 95\\
13 & 25 & 8 & 2296 & 2.46 & $0.96 \pm 0.08$ & 106\\
\end{tabular}
   \caption{Supermoir\'e unit cells for tBLG and graphene/hBN sub-moir\'e lattices with
   comparatively small areas, $l$ is the number of tBLG moir\'e unit cells, $m$ the number of graphene/hBN unit cells, and $p$ the number of selected AA sites in the supermoir\'e unit cell. The column $A_\mathrm{sm}$ refers to the total area of the supermoir\'e unit cell\footnote{For the case $(l=12, m = 13, p = 8)$, the superlattice wavelength of the graphene/hBN moir\'e lattice yields a numerical value of $\lambda_\mathrm{gr/hBN} = (14.47 \pm 0.55)$~nm. The upper bound is thus larger than theoretically possible for a graphene/hBN moir\'e lattice.}.}
    \label{STabular1}
\end{table}
We list the plausible commensurate supermoir\'e structures for a tBLG moir\'e lattice aligned with a graphene/hBN moir\'e lattice in Tab.~\ref{STabular1}. 
As stated in the main manuscript, plausible in this context means that the size of the superstructure is below $50 \, \mathrm{nm} \times 50 \, \mathrm{nm} = 2500 $ nm$^2$ -- otherwise, twist-angle inhomogeneities will probably smear out satellite features. 
The size of the tBLG moir\'e lattice is fixed by the corresponding twist angle of $\theta_\mathrm{tBLG} = 0.987^\circ$. 
While we do not know the relative angle between the graphene layer and the adjacent hBN, it should be close to zero, otherwise the graphene/hBN moir\'e lattice becomes too small.
We search for commensurate supermoir\'e unit cells (Tab.~\ref{STabular1}) by rotating the tBLG moir\'e lattice and the graphene/hBN moir\'e lattice against each other while varying the graphene/hBN twist angle $\theta_{\mathrm{gr/hBN}}$ and thus its size. 
Given the two subtly different origins of the two moir\'e effects (different unit cell size for graphene/hBN, twist between identical lattices for tBLG), the two moir\'e lattices are naturally rotated by $\varphi \approx 90^\circ$ against each other. 
Large deviations from $\varphi = 90^\circ$ --- up  to the sixfold symmetry of the tBLG moir\'e lattice; $\varphi$ is only defined up to $60^{\circ}$ ---  do not correspond to physically realizable structures, as one graphene layer is included in both moir\'e structures.
However, considering reconstructions due to strain, twist-angle modulations and the small moir\'e twist angles, $\varphi $ may vary slightly around this value. \\

The listing in Tab.~\ref{STabular1} contains all commensurate moir\'e structures with $l<15$ and $m < 2l$.
We exclude values $m > 2l$ as the resulting twist angles become larger, $\theta_{\mathrm{gr/hBN}} > 1^\circ$, leading to a short-ranged and quite weak moir\'e potential. 
Furthermore, we exclude larger values of $l$ since in the presence of twist angle disorder, larger supermoir\'e lattice sizes should strongly suppress the visibility of moir\'e satellite peaks. \\

Determining possible filling sequences and values of $p$ purely based on geometry quickly becomes challenging for the larger moir\'e lattices, as the number of AA-sites with slightly different symmetries increases. 
For example, the $l/m = 9/16$ supermoir\'e lattice in Fig.~\ref{FigureS8} features one AA site aligned with an graphene/hBN AA site, 2 tBLG AA sites situated closely to an hBN site, and 6 tBLG AA sites at the center of three graphene/hBN AA sites. 
Their relative degeneracies will depend on details of the two moir\'e lattice potentials.
\newpage

\end{document}